\documentclass[prb, twocolumn]{revtex4}

 \usepackage{amsmath, amssymb}
 \usepackage{bm, color}

\usepackage{graphicx}
\usepackage{hyperref}

\usepackage{natbib}

\newcommand{\be}{\begin{equation}}
\newcommand{\ee}{\end{equation}}

\newcommand{\vrr}{{\bf{r}}}

\newcommand{\vv}{{\bm{v}}}

\newcommand{\vj}{{\bf{j}}}

\newcommand{\vA}{{\bf{A}}}
\newcommand{\vB}{{\bf{B}}}
\newcommand{\vH}{{\bf{H}}}

\newcommand{\si}{\sigma}
\newcommand{\eps}{\epsilon}

\newcommand{\pa}{\partial}

\newcommand{\ti}{\tilde i}
\newcommand{\tR}{\tilde R}
\newcommand{\phio}{| \Phi_0 |}
\newcommand{\ldw}{\ell_{\text{dw}}}
\newcommand{\gdwL}{g_{\text{dw}\circ}}
\newcommand{\gdwS}{g_{\text{dw}|}}

\begin{document}

\title{Stability of topological defects in chiral superconductors: London theory.}

\author{Victor Vakaryuk}
\email[]{vakaryuk@anl.gov}

\affiliation{Materials Science Division, Argonne National Laboratory, Argonne, Illinois 60439, USA}

\begin{abstract}

This paper examines the thermodynamic stability of chiral domain walls and vortices -- topological defects which can exist in chiral superconductors. Using London theory it is demonstrated that at sufficiently small applied and chiral fields the existence of domain walls and vortices in the sample is not favored  and the sample's configuration is a single domain. The particular chirality of the single-domain configuration is neither favored nor disfavored by the applied field. Increasing the field leads to an entry of a domain-wall loop or a vortex into the sample. The formation of a straight domain wall is never preferred in equilibrium. Values of the entry (critical) fields for both types of defects, as well as the equilibrium size of the domain-wall loop, are calculated. We also consider a mesoscopic chiral sample and calculate its zero-field magnetization, susceptibility, and a change in the magnetic moment due to a vortex or a domain wall entry. We show that in the case of a soft domain wall whose  energetics is dominated by the chiral current (and not by the surface tension) its behavior in mesoscopic samples is substantially different from that  in the bulk case and can be used for a controllable transfer of edge excitations. The applicability of these results to $\rm Sr_2RuO_4$ -- a tentative chiral superconductor --  is discussed.

\end{abstract}


\maketitle

\section{Introduction}

Chiral superconductors belong to an exotic class of physical systems whose many-body ground state carries nonzero current and hence breaks time-reversal symmetry. Because the current can assume two time-reversal connected directions,  the ground state of a chiral superconductor is doubly degenerate (chiral). This degeneracy opens a possibility for the existence of extended topological defects -- domain walls -- which connect regions of opposite chirality and exist along with conventional defects such as vortex lines. 

It was recently suggested that graphene at specific doping can support chiral superconductivity.\cite{Honerkamp:Graphene, Nandkishore:2011} Another tentative candidate for a chiral superconductor is $\rm Sr_2RuO_4$ below $1.5\,$K (see Refs.~\onlinecite{MackenzieMaeno:2003, Mineev:2010} for review) which is corroborated by $\mu$SR,\cite{Luke:1998, Luke:2000} the Kerr effect,\cite{Kapitulnik:2006} and phase-sensitive measurements.\cite{Liu:2004, Kidwingira:2006} The candidacy of $\rm Sr_2RuO_4$ is, however, undermined by the fact that  several attempts to detect a surface magnetic field generated by the chiral currents \cite{SQUIDlimits:2005, SQUIDlimits:2007, SQUIDlimits:2010, Kallin:2009} or the magnetic moment associated with them \cite{Jang:2011} have not yielded a positive result.

One of the possible explanations aimed to cut this Gordian knot of seemingly contradicting observations is to assume 
the presence, on a mesoscopic scale, of an alternating chiral domain structure which leads to the substantial field cancellation. Previous studies reported in the literature have focused either on the calculation of the domain wall properties such as surface tension and accompanying chiral current  (see, e.g., Refs.~\onlinecite{Sigrist:1999, Sonin:2009}) or on finding the distribution of magnetic fields assuming a particular domain structure without attempts to justify the latter. \cite{Ichioka:2005, SQUIDlimits:2005, SQUIDlimits:2007, Bluhm:2007, SQUIDlimits:2010}

In this work we use London theory to address the question of the thermodynamic stability of several domain configurations  for a simple sample's geometry (such as cylindrical), where demagnetizing effects can be easily taken into account. We show in particular that in small applied fields the equilibrium domain configuration corresponds to a single-domain sample and that the specific  chirality of this domain is neither favored nor disfavored by the applied field. Upon increasing the applied field either a domain wall loop or a vortex enters the sample; the configuration in which a domain wall forms a straight line which terminates at the edges of the sample is never favored in equilibrium. Values of the entry (critical) fields   as well as of the domain wall loop size are given in terms of model parameters such as the magnitude of the chiral current and a domain wall surface tension. 

Motivated by recent  cantilever magnetometry measurements in $\rm Sr_2RuO_4$ (Ref.~\onlinecite{Jang:2011}), we also consider a mesoscopic chiral sample with a hole for which we calculate  zero-field magnetic moment, susceptibility, and a magnetic moment change due to a vortex or domain wall entry. 
We demonstrate that in the mesoscopic limit the size of a loop formed by a very soft domain wall can be controlled with the applied field and speculate that such an effect can give rise to  a controllable transfer of edge excitations such as Majorana modes.

The paper is organized in the following way. In Section \ref{domain wall description} we give a phenomenological description of a chiral domain wall used throughout the paper. In Section \ref{Gibbs potential} we derive an expression for the Gibbs energy of a chiral superconductor with topological defects in the strong type II limit.  In Section \ref{macro} we apply results of the previous section  to several domain configurations in macroscopic samples. This is the core section of the paper. In Section \ref{micro} we focus on a mesoscopic  geometry. Section \ref{conclusions} is devoted to overall conclusions. Appendix \ref{appendix 1} contains the solution of the London equation for a two-domain circular cylinder of arbitrary dimensions with a hole. Appendix \ref{appendix 2} contains some useful results involving modified Bessel functions.  

\section{Domain wall description}
\label{domain wall description}

Even in the simplest case of an isotropic chiral superconductor the domain wall structure  can be quite complicated and is in general determined by the interplay between the material -- Ginzburg-Landau -- parameters and the domain wall geometry. To describe chiral domain walls  we use a simplified model  in which it is modeled  as a sheet-like object  with surface tension $\si$ which carries chiral current with linear density $2 i$ and is characterized by a winding number $\ldw$.

The presence of the current along a domain wall  is necessitated by the chiral nature of the state in which the internal orbital motion of Cooper pairs, while 
 compensated in the bulk, produces a nonzero charge current $i$ on a boundary with vacuum and $2i$ on a boundary with another domain.  The presence of such a current leads to a discontinuity of magnetic field (or, rather, magnetic induction) across the domain wall.

A general description of a domain wall requires several winding numbers, which reflects the multicomponent nature of the underlying chiral state.\cite{Ferguson:2011}  The winding number relevant to our model, $\ldw$, has a physical meaning of a net flux (in units of flux quantum, cf.~below) generated by the domain wall's  chiral \emph{and} screening currents  in an infinite superconducting medium.\footnote{Our $\ldw$ is $n_+$ in the notation of Ref.~\onlinecite{Ferguson:2011}.} Defined in this way the winding number $\ldw$ depends on the geometry of the domain wall and, in general, need not be an integer.

The domain wall surface tension $\si$ complements our treatment of topological defects by specifying its intrinsic energy per unit area. Although existing calculations seem to indicate that for a chiral $p+ip$ superconductor  $\si >0$,  the author is not aware of a general proof that would exclude the opposite.

We  consider a model in which both the sample and the domain wall are translationally invariant along the direction of the applied field  and focus only on two domain wall configurations -- a straight line and a  circle -- which, due to their high symmetry, admit  straightforward analytical treatment. The main difference between the two configurations is that whereas a circular domain wall creates a nonzero net flux ($\ldw \neq 0$) the net flux created by  a straight domain wall vanishes ($\ldw=0$). It should also be pointed out that both $\si$ and $i$ will in general be different for the two configurations; we do not indicate such a difference explicitly, unless otherwise stated.

We note in passing that the domain structure in  chiral superconductors need not be similar to that in ferromagnetic materials because the latter do not exhibit field screening. 


\section{Surface representation of the Gibbs potential for a chiral superconductor}

\label{Gibbs potential}

Let us start by considering a  superconducting sample placed in a uniform external magnetic field. The distribution of currents and fields in the sample is a function of the applied field and, in thermal equilibrium, can be found through minimization of the corresponding Gibbs potential,  defined as \footnote{Several alternative definitions of the Gibbs potential can be found in the literature. They differ from the one used here by either subtracting the field energy in the absence of a superconductor $-\frac 1 {8\pi} \int \vH_a^2$ and extending the volume integration to the whole space or by including the field energy of the space occupied by the superconductor $\frac 1 {8\pi} \int_{\textsc{sc}} \vB^2$ into the definition of $F_s$.} 
\be
	G= F_s  + \frac 1 {8\pi} \int \!\!d^3 r \,\, (\vB^2 - 2 \vB \cdot \vH),
	\label{def of G}
\ee
where $\vH$ and $\vB$ are magnetic field and induction respectively; the volume integration extends over the space occupied by the superconductor \emph{and} over any cavities contained in it.   The free energy $F_s$ of the sample, which by our definition excludes the field energy given by the $\vB^2$ term in eqn.~(\ref{def of G}), may contain terms describing kinetic energies of charge and spin currents,\cite{Chung:2007} spin-orbit interaction energy,\cite{Miyake:2010} effects of kinematic spin polarization,\cite{Vakaryuk:2009} etc.

In London theory the free energy of a superconductor is approximated by the kinetic energy of supercurrents described by the superfluid velocity $\vv_s$. The sum of the kinetic energy of supercurrents  and the magnetic  field energy  can be written in the following  form:\cite{deGennes}
\begin{widetext}
\be
	\int_{\text{sc}} \!\! \!d^3 r \, \!\!  \left(   \frac 1 2  \rho_s \vv_s^2(\vrr) \! + \! \frac 1 {8\pi}  \vB^2\right) \!\!
	= \!
	 - \frac \phio {16 \pi^2}\!\!  \oint \! d^2 \bm s \cdot (\vB \times  \bm \nabla \theta ) 
	-\frac 1 {8 \pi} \! \oint \!d^2  \bm s \cdot \big( \vB \times \vA \big),
	\label{Tsgeneralfinal}
\ee
\end{widetext}
where $\Phi_0 \equiv hc/2e$ ($<0$), $\vA$ is the vector potential, $\rho_s$ is the superfluid density and $\theta$ is the phase of the order parameter. The volume integration extends over the region of space occupied by the superconductor and $\oint d^2 \bm s $ denotes   the integration over its surface. The above result,  derived under the main assumption of  the London approximation -- uniform superfluid density\footnote{It is straightforward to generalize this result for situations in which the superfluid density is a steplike function of coordinates.} -- is valid for a superconductor of an arbitrary geometry  and relies only on the use of Maxwell's equations and Gauss's theorem. The convenience of such a representation relates to the fact that for relevant geometries the surface integration is usually more straightforward to perform than the volume one.  Moreover, using eqn.~(\ref{Tsgeneralfinal}) one can avoid direct calculation of the magnetic field contribution which is usually quite cumbersome.

Eqn.~(\ref{Tsgeneralfinal}) can also be used in the presence of topological defects if they are treated in the following way: The volume integrals should exclude regions of nonuniform superfluid density associated with the defects, while the surface integrals should be complemented by an integration over the surface that encloses the excluded volume.  The ``missing'' contribution to the free energy can be accounted for by introducing defects' surface energy $F_\si$, which  can be computed from a more general description (e.g., Ginzburg-Landau theory). Such an approach leads to the following representation of the free energy\footnote{We drop the ``potential energy'' term $V(|\psi|)$ from the consideration since in our linear theory contribution of such term away from defects reduces to  a field- and current-independent constant. }
\be
	F_s 
	=
	   F_\si
	  +
	\int_{\text{sc$-$d}} \!\! \!\! \!\!\! d^3 r \,\, \,  \frac 1 2 \, \rho_s \vv_s^2(\vrr),
	\label{def of Fs}
\ee
where the integral now excludes regions with nonuniform superfluid density associated with the defects. In the extreme type II limit $\lambda/ \xi \gg 1$, this approach should give a quantitatively  good approximation for the Ginzburg-Landau energy of a superconductor with topological defects, whereas in the marginal case  $\lambda \gtrsim \xi$ one might hope to get a qualitatively  reasonable description.


\subsection{Gibbs potential of a two-domain cylinder}

For a general sample's geometry,  evaluation of the surface integrals in representation (\ref{Tsgeneralfinal}) is complicated by the spatial  dependence of the magnetic  field. One exception is a geometry that has a translational symmetry along the direction of the applied field (i.e., a cylinder with an \emph{arbitrary} cross section). In this geometry the value of the field on the sample's surface is constant, which can be seen from the application of Amp\`ere's law to a rectangular contour with a side  parallel to the field.

Motivated by recent cantilever magnetometry measurements on mesoscopic annular $\rm Sr_2RuO_4$ samples,\cite{Jang:2011}  we consider a circular hollow cylinder with an onionlike domain structure shown in Fig.~1. The hole which is characterized by 
an integer winding number $\ell_s$ provides, for small applied fields, the only place where vortices can reside,\footnote{The possibility of  a half-quantum vortex state for which $\ell_s$ is a half-integer can be straightforwardly incorporated in the formalism but is ignored here for the sake of simplicity.}  and  the two-domain configuration is the simplest one in which a reduction of the total magnetic moment can be achieved (as observed in  Ref.~\onlinecite{Jang:2011}).

Results of this section can also be used for cylindrical samples with an arbitrarily shaped cross section, provided the distance between the defects and the boundary is much larger than $\lambda$. In this limit, as shown in Section \ref{macro},  a circular domain wall configuration is favored energetically over a configuration in which a straight domain wall runs across the sample and terminates on the sample's boundaries. 

\begin{figure}
      \centering
      \includegraphics[scale=0.4]{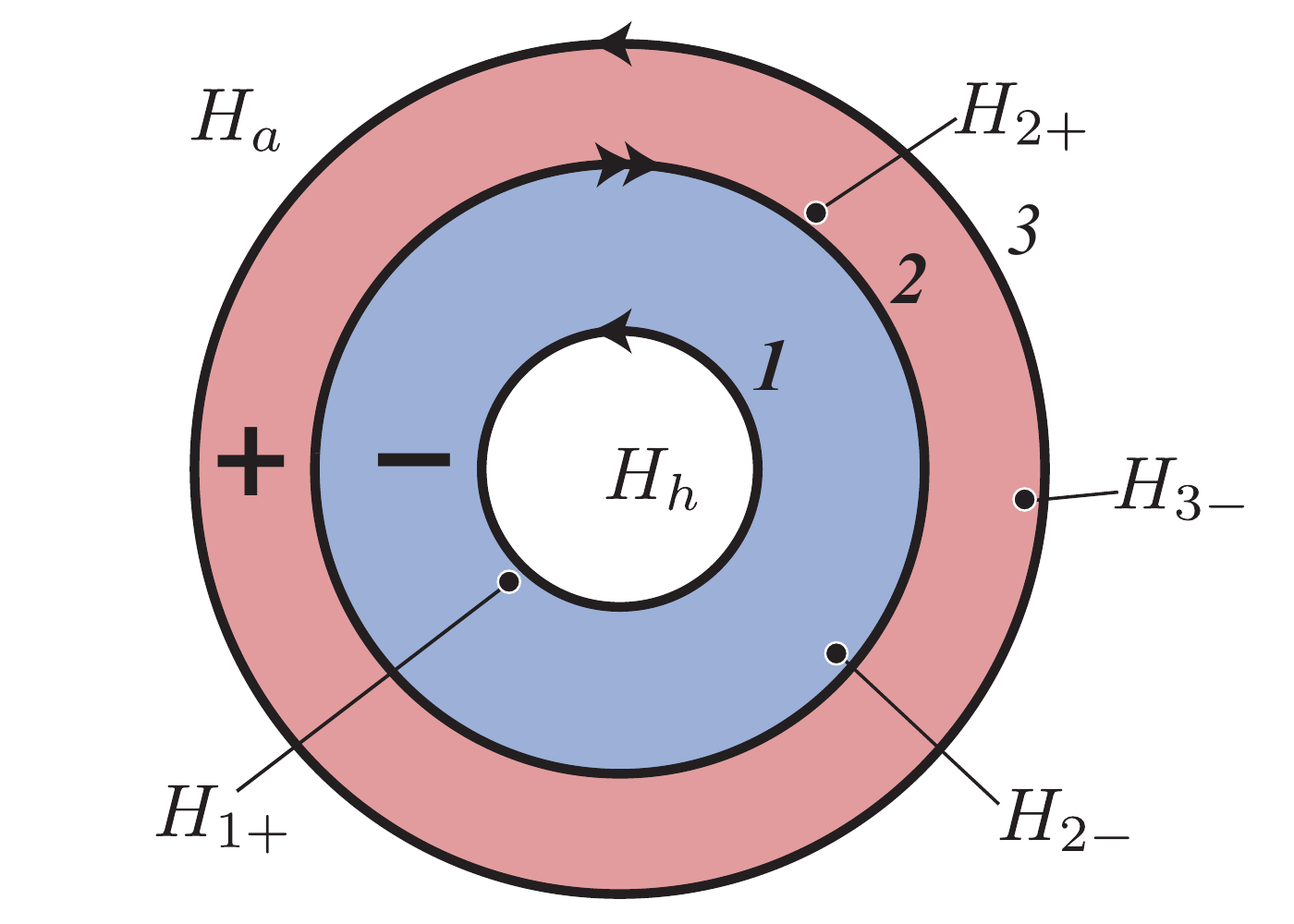}
      \caption{\small Onionlike two-domain configuration of a chiral sample used in the calculation of the Gibbs potential (\ref{g onion}). Arrows indicate the direction of the chiral currents.}
\label{fig1}
\end{figure} 
For a cylindrical geometry 
with the axis parallel to the applied field $\vH_a$  we have $\vB =\vH$ and expression (\ref{def of G}) simplifies to
\be
	G= F_s  + \frac 1 {8\pi} \int \!\!d^3 r \,\, (\vH^2 - 2 \vH \cdot \vH_a),
\label{def G cylinder}
\ee
To make use of eqn.~(\ref{def G cylinder}) we notice that the onionlike geometry shown in Fig.~\ref{fig1} consists of three surfaces: 1 -- the inner surface of the sample, 2 -- the domain wall surface, and 3 -- the outer surface of the sample. The  surfaces are characterized by their respective radii $R_j$. For the domain chirality shown in Fig.~\ref{fig1} surfaces 1 and 3 carry counterclockwise chiral current $i$ while the domain wall 2 carries a clockwise current $2i$. 

Owing to the presence of the chiral current, the magnetic field across each  surface is discontinuous. Let us denote  the field values on the inner ($-$) and outer ($+$) sides of the surface $j$ as $H_{j\pm}$. In this notation $H_{3+}$ and $H_{1-}$ are equivalent to the applied field $H_a$ and to the field in the hole $H_h$, respectively, and the field discontinuities are given by
\be
	H_{3-} - H_a = \ti, \quad H_{2+}-H_{2-} = 2 \ti, \quad H_h - H_{1+} = \ti,
\ee
where we introduced the field jump $\ti \equiv 4 \pi i /c$ which is analogous to a domain wall magnetization used, for example, in Refs.~\onlinecite{Bluhm:2007, Sonin:2009}.

Let $\Phi_j$ denote the total flux through the area limited by the surface $j$ and $\Phi_{ij} \equiv \Phi_i - \Phi_j$. Then, using the definition of the Gibbs potential (\ref{def G cylinder}) and the surface representation of the free energy (\ref{def of Fs}) given by eqn.~(\ref{Tsgeneralfinal}),  the Gibbs potential of this configuration  is given by 
\begin{eqnarray}
	8 \pi g &=& 8\pi f_\si 
	-
	  \phio \ell_s (H_h - H_a) 
	  \label{g onion}
	  \\
	  &-& \phio \ldw (H_{2-} -H_a +\ti )
	  - \ti (\Phi_{21} - \Phi_{32}) - H_a \Phi_3,
	  \nonumber
\end{eqnarray}
where  $g$ and $f$ indicate that energies are taken per unit length in the direction of the applied field. As discussed in Section \ref{domain wall description}, $\ldw$ is a measure of the flux carried by a domain wall in an infinitely large sample and $f_\si$ is the surface energy of the domain wall per unit length.

In  the absence of chiral currents and domain walls, i.e., when both $\ti$ and $\ldw$ are set to zero,
expression (\ref{g onion}) coincides, up to an additive constant\footnote{Cf.~the footnote on different definitions of $G$.}, with that obtained in Ref.~\onlinecite{Arutunian:1983} for a hollow nonchiral cylinder. Notice that one should not expect $G$ to be of a simple form $G\propto \bm M \cdot \bm H_a$ where $\bm M$ is the magnetic moment of the cylinder since, in general, in the absence of the applied field $\bm M \neq 0$.

The application of eqn.~(\ref{g onion}) requires knowledge of fields and fluxes in the system in terms of the applied field and parameters $\ell_s$, $\ldw$, $\ti$, and $R_j$. Such knowledge can be obtained from a solution of the London equation. The general solution of the London equation for an onionlike geometry is given in Appendix~\ref{appendix 1}. We now proceed to the analysis of the ``macroscopic'' limit of this solution, where our results have simple analytical form. In  Section~\ref{micro} we relax the ``macroscopic'' constraint and  consider this  geometry for a sample with arbitrary dimensions.


\section{Stability of topological defects in macroscopic  limit}
\label{macro}

\begin{figure*}
\centering
\includegraphics[scale=0.4]{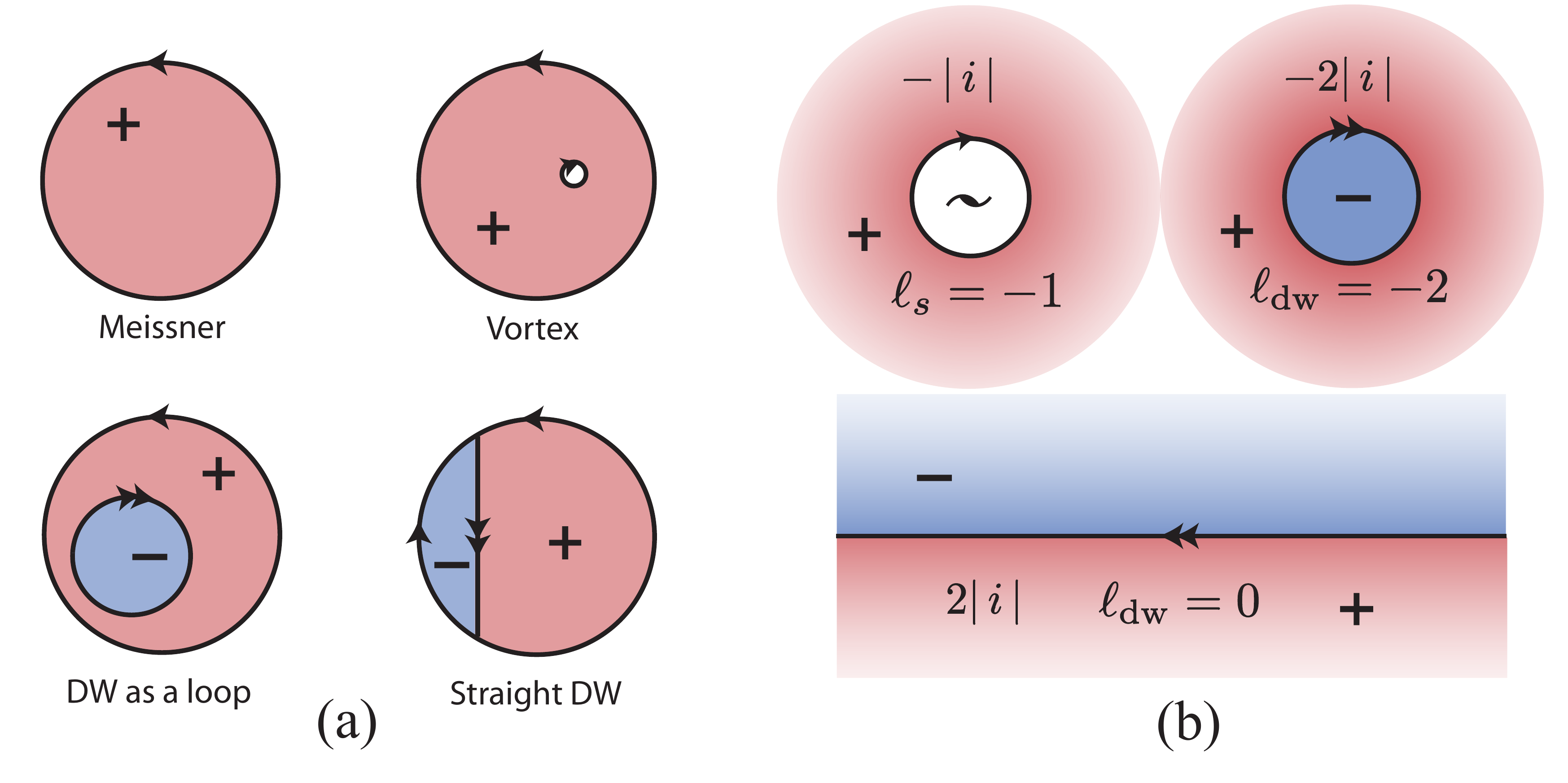}
\caption{\label{fig2} {\small (a) Configurations of the defects considered in Section \ref{macro}. The positive direction of the applied field is out of the page toward the reader. (b) Values of winding numbers and chiral currents for  vortex, loop and  straight domain wall (DW) configurations. The size of the vortex core is exaggerated  for visual purposes.}
}
\end{figure*}

In this section we calculate the Gibbs potential  in the macroscopic  limit when all relevant distances, such as the size of the sample, the distance between defects, and the sample's boundary are larger than $\lambda$. In this limit the precise shape of the boundary and the defect's location relative to it are irrelevant. We consider the four defect configurations shown in Fig.~\ref{fig2}(a): the Meissner state (no defects), the state with a vortex, and the states with a domain wall shaped as a loop or as a straight line. The first three configurations can be obtained as limiting cases of the onionlike geometry of Fig.~\ref{fig1} and hence are described by eqn.~(\ref{g onion}). In the macroscopic, limit Gibbs potentials of  more complicated defect configurations such as combinations of those mentioned above can be written in a similar way.

\subsection{Meissner state}

\label{meissner}

The Meissner state of a cylinder corresponds to a single-domain configuration with no trapped defects, Fig.~\ref{fig2}(a). Its Gibbs potential is obtained from eqn.~(\ref{g onion}) by setting $f_\si$, $\ell_s$, $\ldw$, $R_2$, and $R_1$ to zero  and is given by the expression
\be
	8\pi g_{\rm M} = (\ti - H_a) \Phi_{\rm M},
\ee
where $\Phi_{\rm M} \equiv \Phi_3$ is the net flux through the sample generated by both chiral and screening currents. In the macroscopic limit for a sample of circumference $P$, the Meissner flux $\Phi_{\rm M}$ is given by a plausible expression:
\be
	\Phi_{\rm M} = \lambda P (H_a+\ti),
	\label{PhiM}
\ee
which is a direct consequence of the fact  that both the applied field and the field created by the chiral current are screened over a region of thickness $\lambda$ around the outer edge of the sample.\footnote{Expression (\ref{PhiM}) for the net Meissner flux holds for both solid and hollow samples; in the latter case it is required that  the distance between the hole and the external boundary is much larger than $\lambda$.}
Given expression (\ref{PhiM}), the Gibbs potential of a solid macroscopic cylinder in the Meissner state takes the following form:
\be
	8\pi g_{\rm M}= \lambda P (\ti^2 - H_a^2).
	\label{gM}
\ee
This at first sight counterintuitive result implies
that a chirality of a single-domain Meissner state specified by the sign of $\ti$ is neither favored nor disfavored by an external field. Although proven in the macroscopic limit, this statement is in fact independent of the size of the cylinder and can be shown to hold even when the screening is geometrically limited as in the mesoscopic settings  considered in Section \ref{micro}. 

While it is natural to expect the invariance under the full time-reversal operation, which in the case of the Meissner state involves the reversal of \emph{both} chiral current $\ti$ and the applied field $H_a$, the invariance under the reversal of \emph{either} $\ti$ or $H_a$ alone (partial time-reversal operation) as demonstrated by (\ref{gM}) might be considered a surprising feature.\footnote{Such invariance also holds for a state with a straight domain wall discussed in Section \ref{straight domain wall}  but is violated for a state with a domain wall loop, Section \ref{macro domain wall}.}  
An intriguing question is whether this feature is just a peculiarity of the cylindrical geometry which possesses translational invariance along the direction of the applied field or has a broader validity. While the author does not have a  proof of the latter, a plausibility argument can be given that suggests that the invariance under the partial time-reversal operation can be expected if the sample has a mirror symmetry in the plane perpendicular to the field.

We also note that, as evident from eqn.~(\ref{gM}), chiral currents give a positive contribution to the electromagnetic energy of the system. This statement should also hold for samples with dimensions of the order of or smaller than $\lambda$. For such mesoscopic samples positive chiral contribution to the electromagnetic energy may become comparable with the negative condensation energy whose scale is set by the thermodynamic critical field.  This mechanism  may hinder the formation of the chiral superconducting state and has to be borne in mind when considering the possibility of a chiral pairing in very small samples. \cite{Huo:2011}

\subsection{Vortex state}

\label{vortex}

A vortex state corresponds to a single-domain configuration with a hole with nonzero phase winding $\ell_s$ around it. Such a configuration is obtained from an onionlike geometry  by setting $\ldw=0$, $R_1=\xi \ll \lambda$, and then taking the limit $R_2\to R_1$. Using eqn.~(\ref{g onion}) the Gibbs potential of the vortex state relative to that of the Meissner state is given by the expression
\be
	8\pi (g_{v} - g_{\rm M}) =  - \phio \ell_s (H_v- 2 H_a),
	\label{gv}
\ee 
where $H_v$ is the value of the magnetic  field on the outer side of the surface which defines the normal vortex core; in the notation used in eqn.~(\ref{g onion}), $H_v$ corresponds to $H_{2+}$ after taking the limit $R_2\to R_1$.  In deriving eqn.~(\ref{gv}) we have neglected the vortex core energy and the flux carried by it -- a step which is well justified in the extreme type II limit used here.


For large applied fields, the energy difference (\ref{gv}) is negative, which means that the Meissner state is thermodynamically unstable. The critical field for the vortex entry  is determined by the following equation:
\be
	H_{c1,v}=H_{v}/2.
\ee
The field $H_v$ can be found by either solving the London equation in the macroscopic limit or by taking appropriate limits in the results for the onionlike geometry given in the Appendix \ref{appendix 1}. In doing so one obtains
\be
	H_{c1,v} = - \frac{\phio \ell_s}{4 \pi \lambda^2} \ln (2 \lambda/\xi).
	\label{Hc1 vortex}
\ee
Thus, the value of the first critical field for a vortex entry  in a chiral superconductor is independent of both the magnitude and the sign  on the chiral current $\ti$ and, within logarithmic precision, coincides with $H_{c1}$ for nonchiral superconductors\footnote{We use convention in which the electron's charge $e$ is negative and the phase $\theta$ of the order parameter is defined as $\psi = |\psi| \exp +i\theta$. This implies that a gauge-invariant phase is given by the following combination $\bm \nabla \theta + 2\pi \vA /\phio$ where $\vA$ is the vector potential such  that for a positive applied flux the equilibrium winding number $\ell_s$ is negative. } (see, e.g., Ref.~\onlinecite{Shmidt:1974}).  Alternatively, for a fixed chirality, eqn.~(\ref{Hc1 vortex}) demonstrates the absence of the field-reversal splitting of $H_{c1,v}$. It should be noted, however, that inclusion of the vortex core energy in (\ref{gv}), which tends to be chirality-dependent,\cite{Hc1FieldReversalSplitting} will result in nonzero field-reversal splitting of $H_{c1,v}$.

One might wonder what happens if the magnitude of the chiral current $\ti$ is such that the magnetic field created in its immediate neighborhood is larger than $H_{c1,v}$. While this question cannot be answered within the macroscopic approximation used in this section, eqn.~(\ref{Hc1 vortex}) suggests that, for applied fields smaller than $H_{c1,v}$, vortices generated by the chiral current's magnetic field will tend to stay away from the bulk   ``decorating'' edges of the sample and domain boundaries (if present).

Another conclusion that can be drawn from the results of this section is that, in thermal equilibrium in the absence of the applied field, the total flux trapped by a chiral cavity, located at distance much larger than $\lambda$ away from the sample's boundary, is zero. This follows from setting $H_a=0$ in expression (\ref{gv}) and then minimizing it with respect to $\ell_s$.

\subsection{State with a domain wall loop}

\label{macro domain wall}

We now proceed to the configuration in which a domain wall forms a circular  loop,  i.e., terminates in the sample forming an ``island'' of opposite chirality. The Gibbs potential $\gdwL$ of such a configuration is obtained from eqn.~(\ref{g onion}) by setting $\ell_s=0$ and taking the limit $R_1 \to 0$. This leads to the  expression
\begin{eqnarray}
	8\pi \gdwL
	&=&
	8 \pi f_\si
	-\phio\ldw (H_{2-} - H_a + \ti)
	\nonumber
	\\
	&-&
	2\ti \Phi_2
	+
	\Phi_3(\ti - H_a),
	\label{def of G for dw}
\end{eqnarray}
where $f_\si$ is the surface energy defined after eqn.~(\ref{g onion}). $H_{2-}$ is the field on the inner side of the domain's boundary, $\Phi_2$ is the flux through the area limited by it, and $\Phi_3$ is the total flux through the sample, which includes the screening contribution and the flux created by the domain. In the macroscopic limit these quantities can be found by taking the appropriate limit in the general solution for the field distribution,
 which is given in Appendix \ref{appendix 1}. In this way we obtain
\be
\begin{split}	
	H_{2-} &=- \frac{\phio \ldw} {4\pi R \lambda} - \ti,
	\\
	\Phi_2 & =- \frac 1 2 \phio \ldw - \ti 2\pi R \lambda,
	\\
	\Phi_3 & = \Phi_{\rm M} -\ldw \phio,
\end{split}
\ee
where $R$ is the radius of the domain island [$\equiv R_2$ in the notation of eqn.~(\ref{g onion})].  The flux $\Phi_3$ consists of the flux $\Phi_{\rm M}$  generated by the boundary of the sample [cf.~eqn.~({\ref{PhiM})] and of the flux carried by a domain wall loop, $\ldw \Phi_0$. Plugging these results into eqn.~(\ref{def of G for dw}) yields the following expression for the Gibbs potential of the circular domain wall configuration:
\begin{eqnarray}
	8\pi(\gdwL - g_{\rm M}) &=& \frac{\phio^2 \ldw^2}{4\pi R \lambda} + (\ti^2 + 4\pi \si/\lambda) \, 4\pi R \lambda 
	\nonumber
	\\
	&+& 
	2 \phio \ldw H_a,
	\label{gdwL}
\end{eqnarray}
where $g_{\rm M}$ is the Gibbs potential of the Meissner state and $\si$ is the surface tension of the domain wall, such that $f_\si = 2\pi R \si$.  We first minimize the above expression  with respect to $R$ which yields the equilibrium size of the domain island:
\be
	R_\circ = \frac{\phio |\ldw|}{4\pi \lambda \sqrt{\ti^2 + 4\pi \si /\lambda}}.
	\label{R loop}
\ee
Evaluated at $R_\circ$ the difference (\ref{gdwL}) is positive for small applied fields; upon increasing the field the difference (\ref{gdwL}) becomes negative at some value of $H_a$ which defines the critical field for the creation of a domain wall loop. In other words, circular domains such as that shown in Fig.~\ref{fig2}(a) will be thermodynamically stable only if the applied field $H_a$ exceeds a critical value $H_{c1,\text{dw}}$, defined by
\be
	H_{c1,\text{dw}} = \sqrt{\ti^2 + 4\pi \si/\lambda}.
	\label{Hc1 dw}
\ee
Notice that unlike $H_{c1,v}$ for a vortex entry, eqn.~(\ref{Hc1 vortex}), the critical field for the domain wall loop entry  depends on the chiral current $\ti$ but is independent of the flux carried by the defect. The reason for the latter is the $\ldw$ dependence of the equilibrium domain size as specified by eqn.~(\ref{R loop}).



Let $+-+$ denote the chirality arrangement of a domain wall loop state shown in Fig.~\ref{fig2}(a). Unlike the Meissner state discussed in Section \ref{meissner}, the energy of this state, eqn.~(\ref{gdwL}), is \emph{not} invariant under reversal of the applied field. Equivalently, for a fixed applied field the energies of the $+-+$ arrangement and of its time-reversal counterpart $-+-$ (obtained by changing the direction of the chiral currents and the sign of $\ldw$) are different. In particular, for a positive applied field the energy of the $-+-$ arrangement is larger than that of $+-+$. Although the energy of the former state can be lowered by adding vortices, it will still be larger than either the $+-+$ arrangement or a pure vortex state and hence cannot correspond to a true equilibrium. 





Putting together eqns.~(\ref{Hc1 vortex}, \ref{R loop}, and \ref{Hc1 dw}) leads to the following expression for the size of the domain island:
\be
	R_\circ /\lambda \propto H_{c1,v}/H_{c1,\text{dw}},
\ee
i.e., $R_\circ$ scales as the ratio of critical fields of the vortex and domain entries. Strictly speaking, the formulation that leads to this scaling is valid only if $R_\circ /\lambda \gg 1$. However, because of the exponential falloff of the screening currents, one might  expect that it is qualitatively correct even in the limiting case of relatively small domains when $R_\circ /\lambda \gtrsim 1$.


To estimate the actual value of $H_{c1,\text{dw}}$ for a given material a knowledge of  chiral current $\ti$ and domain wall surface tension $\si$ is required. These can be calculated using Ginzburg-Landau  theory and  turn out to depend on various material and geometrical parameters  such as Ginzburg-Landau expansion coefficients and the orientation of the domain wall relative to the crystal axes (see, e.g., Refs.~\onlinecite{Sigrist:1999, Sonin:2009}). Ignoring for simplicity material anisotropy, one can conclude that
\be
	4 \pi \si/\lambda  = \eps_1 \frac{\phio^2}{4 \pi^2 \lambda^3 \xi},  \quad \ti^2 = \eps_2 \frac{\phio^2}{4\pi^2 \lambda^4},
	\label{estimates}
\ee
where $\eps_{1,2}$ are dimensionless parameters which in a weak-coupling BCS limit are of the order of 1.\cite{Sigrist:1999, Sonin:2009} We now define the following parameter:
\be
	\varkappa_d \equiv 4\pi \si /(\ti^2 \lambda) = \eps_1 \lambda/( \eps_2 \xi).
	\label{def of kappa}
\ee
As can be seen from eqns.~(\ref{R loop}) and (\ref{Hc1 dw}) this parameter  determines whether the energetics of the domain wall is dominated by the chiral current (``soft'' domain wall, $\varkappa_d \ll 1$) or by the surface tension (``hard'' domain wall, $\varkappa_d \gg 1$). In the weak-coupling limit when $\eps_{1,2} \sim 1$, one generally expects that  $\varkappa_d \approx \lambda/\xi$. Provided the weak-coupling limit is applicable for $\rm Sr_2 RuO_4$ ($\lambda/\xi \sim 1$ and hence $\varkappa_d \sim 1$), one would expect that  $H_{c1,v} \sim H_{c1, \text{dw}}$ and the domain size $R_\circ \sim \lambda$. However, given the unconventional nature of superconductivity in $\rm Sr_2RuO_4$, the applicability of the weak-coupling results to this material remains an open question.

\subsection{State with a straight domain wall that terminates at the edges}

\label{straight domain wall}

A straight domain wall configuration that terminates at the edges of the sample is qualitatively different from the closed configuration discussed earlier. Unlike the latter, the total flux carried by a straight domain wall is zero (see, e.g., Ref.~\onlinecite{Ferguson:2011}), which substantially changes its energetics. Let $\Phi_\pm$ denote the total flux carried by $\pm$ domains shown in Fig.~\ref{fig2}(a). The Gibbs potential of such a state can be found along the lines that led to eqn.~(\ref{g onion}), and is given by
\be
	8 \pi \gdwS =  8\pi f_\si+ \ti (\Phi_+ - \Phi_-)  - H_a (\Phi_+ + \Phi_-).
\ee
In the macroscopic limit, fluxes $\Phi_\pm$ can be easily computed, which leads to the following expression for $\gdwS$:
\be
	8\pi(\gdwS - g_{\rm M})
	=
	2 R \lambda \, (\ti^2  + 4\pi \si / \lambda),
	\label{g straight}
\ee
where $g_{\rm M}$ is the Gibbs potential of the Meissner state and $R$ is the length of the domain wall segment. Provided that the surface energy $\si>0$,  the field-independent expression (\ref{g straight}) is always positive,  which implies that a straight domain wall configuration is thermodynamically unstable relative either  to the Meissner state or to the state with a domain wall loop. 

\begin{figure*}
\centering
\includegraphics[scale=0.6]{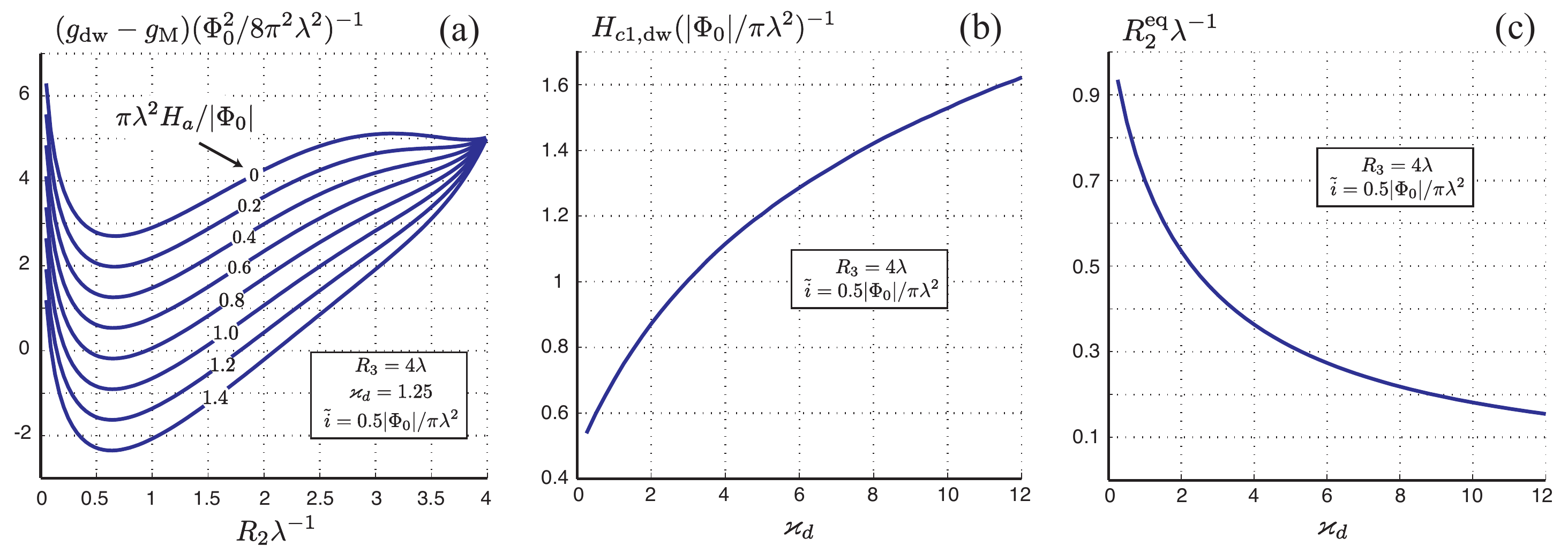}
\caption{\label{fig:defects} {\small (a) The spatial profile of the  Gibbs potential of a mesoscopic chiral cylinder with a domain wall loop as a function of its size for several applied fields. The potential is given relative to that of the Meissner state. Two metastable equilibria are clearly visible at zero applied field. (b,c) Critical field $H_{c1,\rm dw}$ and the equilibrium size $R_2^{\rm eq}$ of the domain wall loop at $H_a=H_{c1,\rm dw}$ as functions of the parameter $\varkappa_d$, defined by eqn.~(\ref{def of kappa}). It is assumed that the variation of $\varkappa_d$ is entirely due to the variation of either $\eps_1$ or $\eps_2$. Surprisingly, even for a mesoscopic sample, values of  $H_{c1,\rm dw}$ and $R_2^{\rm eq}$ shown in (b) and (c) agree quantitatively well with analytical results (\ref{Hc1 dw})  and (\ref{R loop}) obtained for a macroscopic sample. }
}
\label{fig3}
\end{figure*}

\subsection{Summary of Section \ref{macro}}

To summarize, in this section we obtained critical fields for a domain wall $H_{c1, {\rm dw}}$, eqn.~(\ref{Hc1 dw}), and a vortex $H_{c1,v}$, eqn.~(\ref{Hc1 vortex}), entries into a chiral superconductor in the macroscopic limit (all relevant dimensions much larger than $\lambda$). 
It was shown  that the preferred domain wall configuration is that of a loop whose equilibrium size is given by eqn.~(\ref{R loop}). These results imply, in particular, that, at fields above $H_{c1, {\rm dw}}$, a cross section of the domain structure of a macroscopic sample in the direction perpendicular to the field is that of a plum pudding -- a single domain populated by domain islands of opposite chirality.

It was also pointed out that a statement, often encountered in the literature, that  cooling in the field should reduce domains by biasing the system to one chirality (field training) does not refer to the thermodynamic equilibrium. This can be seen from the eqn.~(\ref{gM}) for the Gibbs potential of a single-domain sample that  does not contain terms linear in the chiral current and hence cannot differentiate between  domains of opposite chirality. Upon increasing the field a topological defect which corresponds to the minimal of the two fields $H_{c1,v}$ and $H_{c1,\text{dw}}$ enters the sample and for large fields both vortices and domain wall loops will be present.

However, even in thermal equilibrium,  the direction of the applied field can affect relative chiralities;  for example, for a positive field the $+-+$ domain wall loop configuration of Fig.~\ref{fig2}(a) is favored over its time reversal $-+-$.

\section{Magnetic response of a mesoscopic chiral sample}

\label{micro}

In this section we consider the thermodynamic stability of domain walls and vortices in mesoscopic chiral samples whose relevant dimensions are comparable to $\lambda$.\footnote{Assuming that the chiral superconductivity is possible in mesoscopic samples. Cf.~discussion at the end of Section~\ref{meissner}.}  This problem is motivated by recent cantilever magnetometry measurements done on small $\rm Sr_2RuO_4$ particles. \cite{Jang:2011} Although the main aim of Ref.~\onlinecite{Jang:2011} was to probe the existence of half-quantum vortices, it is interesting to examine whether the observations reported there shed any light on the question of the chiral nature of $\rm Sr_2 RuO_4$.

As in Section \ref{macro} we will make use of expression (\ref{g onion}) to evaluate the Gibbs potential for the onionlike geometry shown in Fig.~\ref{fig1}. It is  assumed that vortices  present in the system reside only in the cylinder's hole, which guarantees  a contour-independent definition of  the vortex winding number $\ell_s$. This assumption excludes the possibility of wall vortices and limits our consideration to relatively small applied fields and moderate chiral currents (cf.~discussion at the end of Section~\ref{vortex}). Notice, however, that, as demonstrated in Ref.~\onlinecite{Jang:2011}, in a confined geometry with geometrically reduced screening the field required for the wall vortex entry can be substantially larger than the bulk $H_{c1,v}$ [given by eqn.~(\ref{Hc1 vortex})].

It is convenient to introducing the notation
\be
\begin{split}
	a_{jk} = K_0(j) I_0(k) - I_0(j) K_0 (k),
	\\
	b_{jk} \equiv K_2(j) I_0 (k) - I_2(j) K_0 (k),
\label{def of a and b}
\end{split}
\ee
where $I_n$ and $K_n$ are modified Bessel functions of $n$-th order and $K_0(j)\equiv K_0(R_j/\lambda)$, etc.; $R_1$, $R_3$, and $R_2$  stand for the radii of the inner and outer surfaces and for the radius of the circular domain wall loop, respectively (see Fig.~\ref{fig1}).

The magnetic moment of the cylinder is given by the following expression:
\be
	M = L(\Phi_3 - \pi R_3^2 H_a)/ 4\pi,
\ee
where $L$ is the height of the cylinder and $\Phi_3$ is the total flux through the area limited by outer boundary 3.  Unlike the macroscopic limit, the values of fields and fluxes required to evaluate quantities of interest are no longer given by simple analytical expressions and are relegated to Appendix \ref{appendix 1}, where the general solution of the London equation for this geometry is presented.

In the zero applied field the ground state of the system is obtained  through minimization of (\ref{g onion}) and corresponds to   $\ldw=0$, $\ell_s=0$ i.e.~to a single-domain defect-free state. The zero-field magnetic moment $M_0$ of a single-domain chiral cylinder is then equal to
\be
	M_0= (\ti L/4\pi) \big( \pi R_3^2 - 2 \pi \lambda^2/b_{13} \big)
	+
	\ti \chi_m,
	\label{def of M0}
\ee
where $\chi_m$ is the magnetic susceptibility,  $\chi_m = \pa M / \pa H_a$. The magnetic susceptibility is determined by  the system's dimensions $R_1$ and $R_3$ and does not depend on the chiral current $\ti$:
\be
	4\pi L^{-1} \chi_m 
	=
	\frac{\pi R_3^2}{a_{13}}
	\,
	\left(
	{b_{31}} - \frac {4 \lambda^4}{R_1^2 R_3^2} \, \frac 1 {b_{13}}
	\right).
	\label{def of chim}
\ee
It is interesting to note that the result (\ref{def of chim}) also holds  for \emph{nonzero} $\ell_s$ and $\ldw$, i.e., in the presence of either (hole) vortices or a domain wall loop as long as the radius of the latter is field independent. In mesoscopic settings the independence of equilibrium $R_2$ on $H_a$ can be expected for a ``hard'' domain wall whose energetics is dominated by the surface tension $\si$. In the opposite limit of a ``soft'' domain wall whose behavior  is dominated by the chiral current $\ti$ and not by the surface tension $\si$ one can expect significant variations of $R_2$ with $H_a$, as demonstrated below. Such variations lead to a deviation of the response from the simple linear form described by (\ref{def of chim}).

Upon increasing the applied field the system undergoes a transition into a state in which either $\ell_s$ or $\ldw$ is nonzero. The ordering of these events can be determined from the comparison of the critical fields required for the entry of the defects. The critical field for a hole vortex entry is given by
\be
	H_{c1,v} = \frac{\phio} {2 \pi R_1^2} \, \frac {a_{13}} {b_{13} - 2 \lambda^2/R_1^2}.
	\label{meso Hc1v}
\ee
Setting $R_1 \to \xi$ and $R_3 \to \infty$ in the expression above, one recovers the bulk limit given by eqn.~(\ref{Hc1 vortex}). 

To find the critical field for a circular domain wall entry one first needs to know its equilibrium size, which can be found through the minimization  of the Gibbs potential (\ref{g onion}) with respect to $R_2$. Unlike the macroscopic limit (Section \ref{macro domain wall}), the spatial profile of the Gibbs potential in geometries with constrained screening can be quite complicated and may include several metastable equilibria [see Fig.~\ref{fig3}a],  which obstructs  transparent analytical treatment. Numerical results for $H_{c1,\rm dw}$ and $R_{2}^{\rm eq}$ are given in Fig.~\ref{fig3}(b) and (c),  where they are plotted as a function of the parameter $\varkappa_d$ which characterizes the interplay between chiral currents and the surface tension,  eqn.~(\ref{def of kappa}). It has also been checked that the dependencies shown inn Fig.~\ref{fig3}(b) and (c)  also describe a cylinder with a hole, provided that $R_2^{\rm eq}$ is constrained to lie between $R_1$ and $R_3$ and $H_{c1,\rm dw}$ is constrained by the values of $\varkappa_d$ which correspond to $R_2^{\rm eq} = R_1$ and $R_2^{\rm eq} =R_3$.

It is important to emphasize  that both $H_{c1,v}$  and $H_{c1,\rm dw}$ discussed above are computed for a defect-free sample. Only one of these fields has a physical meaning; for example, if it turns out that $H_{c1,v} < H_{c1,\rm dw}$ then the value of the latter needs to be recalculated in the presence of a vortex.

The entry of a defect into the sample leads to a jump in the magnetic moment. Such a jump can be evaluated using the results of Appendix \ref{appendix 1} and is given by
\be
	\Delta M_v =  \frac {\phio L}{4\pi} 
	\Delta \ell_s \left( 1 - \frac{2 \lambda^2}{R_1^2 b_{13}} \right)
\ee
for a hole vortex entry and 
\be
	\Delta M_{\rm dw} =  \frac { \phio  L}{4\pi} 
	\Delta \ldw \left( 1 - \frac{b_{12}}{ b_{13}} \right)
\ee
for a circular domain wall entry. While $\Delta M_v$ is independent of chiral current, $\Delta M_{\rm dw}$ depends on the domain wall size $R_2$, which is determined through the energy minimization and hence implicitly depends on $\ti$. In the limit $R_2 \to R_1$ we have $\Delta M_v/\Delta M_{\rm dw} = \Delta \ell_s /\Delta \ldw$ and if $R_2 \to R_3$ then $\Delta M_{\rm dw} \to 0$.

We now come back to the  case of an extremely soft domain wall mentioned earlier. Fig.~\ref{fig4} shows the radius of a circular domain wall with $\si \to 0$ as a function of the applied field. As the applied field is increased the domain wall moves continuously from the outer to the inner surface of the cylinder.\footnote{Notice that depending on the actual value of $H_{c1,v}$ a higher-field part of the dependence shown in Fig.~\ref{fig4} may correspond to a metastable state.} Given the possibility that chiral boundaries can carry topologically nontrivial excitations such as Majorana modes (see, e.g., Ref.~\onlinecite{Grosfeld:2011}) one may speculate that such a process  can be used to perform a controllable transfer of excitations between the edges of the sample.

\begin{figure}
      \centering
      \includegraphics[scale=0.7]{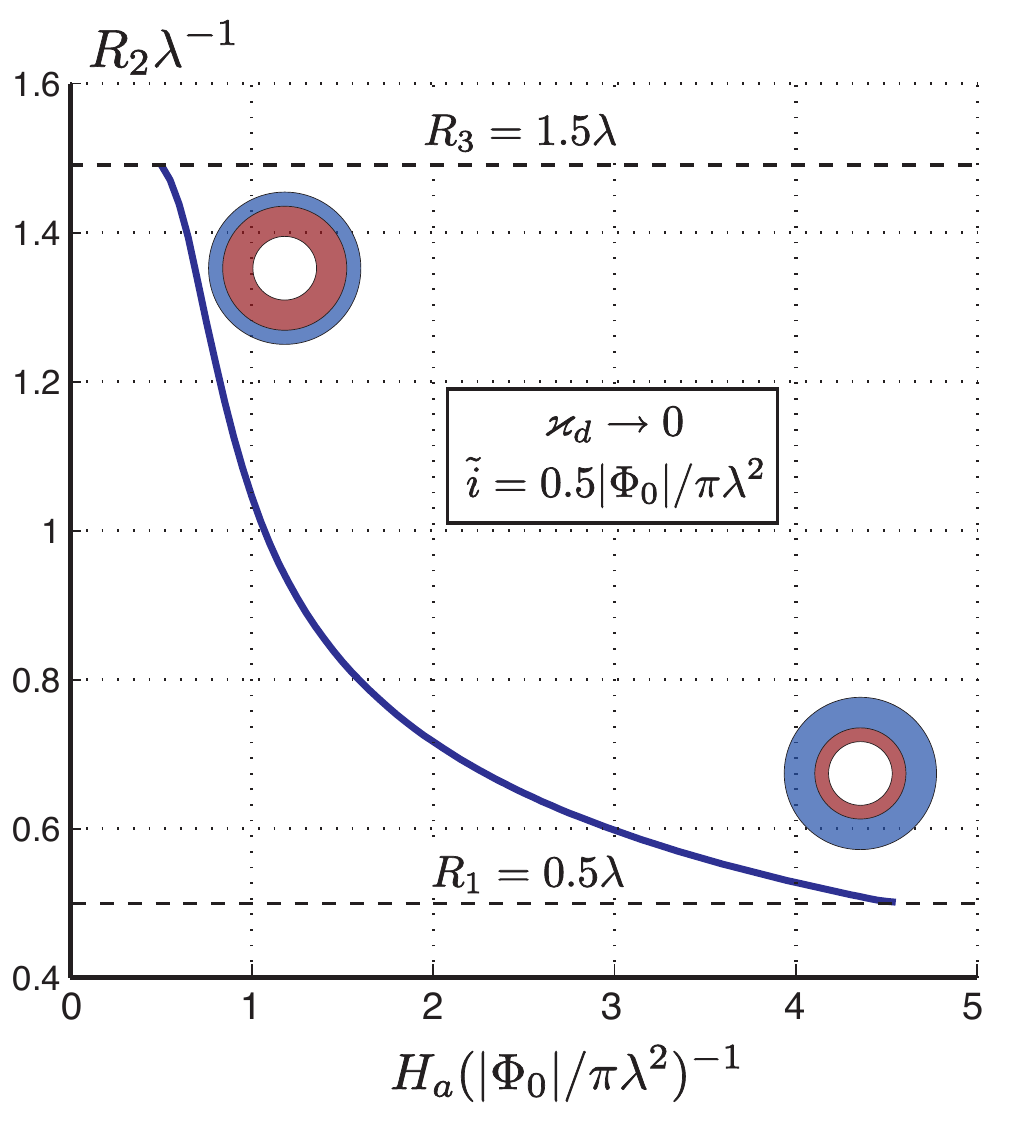}
      \caption{\small Radius of a soft domain wall  loop as a function of the applied field. The demonstrated dependence of $R_2$ on $H_a$  might be useful for performing controllable transfer of edge excitations. }
\label{fig4}
\end{figure} 

\subsection{Application to S\lowercase{r2}R\lowercase{u}O\lowercase{4}}

\label{discussion}

We now turn to the question of the interpretation of the results of Jang \textit{et al}.~\cite{Jang:2011} in terms of possible
 chiral superconductivity.  Jang \textit{et al}.~reported cantilever magnetometry measurements of a mesoscopic $\rm Sr_2RuO_4$ particle with approximate dimensions $R_1=390\,$nm, $R_3=850\,$nm, and $L=350\,$nm;  the magnetic moment sensitivity was of the order of  $10^{-15}\, $e.m.u.\footnote{This is unprecedented moment sensitivity. Compare with the sensitivity of torque magnetization measurements in cuprates which is about $10^{-9}\,$e.m.u., see e.g. L.~Lu \textit{et al}., Phys.~Rev.~B {\bf 81}, 054510 (2010).} The range of fields used in the measurements was such as to cover  the first expected  entry for a hole vortex, eqn.~(\ref{meso Hc1v}).
 
 The quantities $\chi_m$, $\Delta M_v$ and $H_{c1,v}$ computed earlier are independent of chiral current and the only quantity that can be used to estimate $\ti$ independently of $\si$ is $M_0$. Within the noise resolution, measurements reported in Ref.~\onlinecite{Jang:2011} did not observe zero-field moment, which sets the limit $M_0<10^{-15}\,$e.m.u. Using expressions (\ref{def of M0}) and (\ref{def of chim}) and the dimensions of the sample quoted above, we obtain the following upper bound for the magnitude of the chiral current:
\be
	i <10^{-3} \times i_{\rm wc}
	\label{i limit}
\ee
where the weak-coupling value $i_{\rm wc}$ for $\rm Sr_2 RuO_4$ is obtained from eqns.~(\ref{estimates}) by setting $\eps_2=1$, $\lambda =200\,$nm and is approximately equal to $1.9\times 10^{11}\,$\,statamp/cm. Limit (\ref{i limit}) is consistent with scanning SQUID microscopy measurements\cite{SQUIDlimits:2010} where, assuming the domain size of the order of $1\,\mu$, measured $i$ was estimated to be less than 0.1\% of the weak-coupling value.

\section{Conclusions}

\label{conclusions}

Let us review the quantitative results of this paper. We have considered thermodynamic stability of two types of topological defects -- vortices and domain walls -- which can exist in  a chiral superconductor. Using the London theory it was shown that in the zero applied field a macroscopic chiral  sample is either defect-free or has defects which are expelled toward the edges. The first situation is realized if the chiral currents are small and  the second requires them to be sufficiently large such that the  magnitude of the field created  in the immediate neighborhood of the edge chiral current is  larger than  a critical field required for a defect entry.  

It was shown that a preferred configuration of the domain wall is that of a loop; a straight domain wall is never favored in thermodynamic equilibrium. Domain wall loops can exist in the superconducting bulk only if the applied field is larger than $H_{c1,\rm dw}$, which depends on both the magnitude of the chiral current and the surface tension of the domain wall. The critical field required for a bulk vortex entry is not affected by the presence of the chiral currents. 

We have also considered magnetic response and defect stability in mesoscopic chiral samples. It was shown that for a very soft domain wall its size can be controlled by the applied field. This phenomenon can potentially provide a mechanism for a controlled transfer of edge excitations such as Majorana modes.

There are several possible extensions to this work which can be treated in the general framework outlined in Section \ref{Gibbs potential}. An obvious one is to generalize the results presented here for sample geometries that are not translationally invariant along the applied field. In particular, it is interesting to inquire whether the conclusion that the chirality of a single-domain sample is not favored by the applied field holds for other types of geometries. One geometry which seems to be analytically tractable is that of the Pearl limit, in which the thickness of the sample is smaller than the penetration depth. This might be particular relevant in connections with recent speculations about chiral superconductivity in graphene\cite{Honerkamp:Graphene, Nandkishore:2011}.

One can also consider domain structures that are not translationally invariant along the direction of the field. Such a possibility which was suggested in Ref.~\onlinecite{SQUIDlimits:2010}  is very attractive because it has the potential to reconcile the absence of the chiral field in $\rm Sr_2RuO_4$ as observed by scanning measurements with nonzero Kerr and $\mu$SR signals.\footnote{Kerr measurements reported in Ref.~\onlinecite{Kapitulnik:2006} are confined to the surface since the penetration depth of the EM radiation used in Ref.~\onlinecite{Kapitulnik:2006} is of the order of 100\,nm.}

Another possible extension would deal with the interaction between domain wall loops and domain wall loops and vortices. The interaction between such defects may have measurable signatures in the magnetization curves of macroscopic chiral samples.

\section{Acknowledgments}

It is a pleasure to thank  David Ferguson, Tony Leggett  and Catherine Kallin for useful discussions and Valentin Stanev for critical reading of the manuscript. I am also very grateful to  Alex Levchenko and Mike Norman for continuous encouragement and support. The financial support was provided  by the Center for Emergent Superconductivity, an Energy Frontier Research Center funded by the U.S.~DOE, Office of Science, under Award No.~DE-AC0298CH1088.

\appendix

\section{Solution of the London equation for an onion-like geometry}

\label{appendix 1}

In this appendix we consider the solution of the London equation for the onionlike circular two-domain configuration shown in Fig.~\ref{fig1}; a single-domain configuration can be obtained as a limiting case by moving the domain wall to the inner or outer boundary of the sample.

As  mentioned earlier, in the extreme type II limit, the  magnetic properties of a chiral domain wall can be modeled by replacing it with a sheet current $i$ if the domain wall is on the surface and $2i$ if it is in the bulk. The current  carried by e.g. a surface domain wall is $iL$, where $L$ is the length of the domain wall across the direction of the current. In a cylindrical geometry with an external field parallel to the cylinder's axis, a sheet current $i$ is equivalent to a boundary condition for the magnetic field,
\be
	H_{\, \cdot \uparrow} - H_{\uparrow \cdot} = \frac {4\pi}c i \equiv \ti,
\label{def of bc}
\ee
where the arrow in the subscripts indicates the direction of the current and a dot indicates the side at which the field is taken. In the London limit the calculation of the magnetic response of a sample with a given domain structure is thus reduced to solving the London equation in each domain and then matching solutions using appropriate boundary conditions.  Let $R_2$ denote the radius of the domain wall and $R_1$ and $R_3$ be the  inner and outer radii of the cylinder. Using eqn.~(\ref{def of bc}), the boundary conditions for the field can be written in terms of the fields on the domain boundaries:
\be
	H_{3-} = H_a +\ti,
	\quad
	H_{2+}- H_{2-} = 2 \ti,
	\quad
	H_{1+}=H_h - \ti,
	\label{field jumps}
\ee
where $H_{3-} \equiv H(R_3-0)$, $H_{1+} \equiv H(R_1+0)$, $H_{2^\pm}\equiv H(R_2 \pm 0)$ and $H_a$ and $H_h$ are the applied field and the field in the hole respectively. In cylindrical coordinates, the solution of the London equation can be written in terms of Bessel functions $I_0$ and $K_0$:
\be
\begin{split}
	r \in (R_1,R_2):
	\quad
	H(r) = c_{12} I_0(r/\lambda) + c_{12}' K_0(r/\lambda),
	\\
	r \in (R_2,R_3):
	\quad
	H(r) = c_{23} I_0(r/\lambda) + c_{23}' K_0(r/\lambda),
	\label{London eqn. solutions}
\end{split}
\ee
where the constants $c_{jk}$ and $c_{jk}'$ are determined by fields on the domain boundaries,
\be
\begin{split}
	c_{12} = a_{12}^{-1} (H_{2-} K_0(1) - H_{1+} K_0(2) ),
	\\
	c_{12}' = a_{12}^{-1} (H_{1+} I_0(2) - H_{2-} I_0(1) ),
	\\
	c_{23} = a_{23}^{-1} (H_{3-} K_0(2) - H_{2+} K_0(3) ),
	\\
	c_{23}' = a_{23}^{-1} (H_{2+} I_0(3) - H_{3-} I_0(2) ),
	\label{def of c}
\end{split}
\ee
with $a_{jk}$ defined as
\be
	a_{jk} = K_0(j) I_0(k) - I_0(j) K_0 (k),
\label{def of a}
\ee
where $K_0(j)\equiv K_0(R_j)$, etc. 

The equations above determine the field distribution  through yet-unknown values of the fields on the boundaries. To find the latter, one can use additional constraints such as those provided by the Feynman-Onsager (FO) quantization condition,  which is obtained from the   London form of the Ginzburg-Landau equation for the current:
\be
	\vj_s 
	=
	- \frac {c \phio}{8 \pi^2 \lambda^2} \big ( \bm \nabla \theta + \frac {2 \pi}\phio \, \vA \big),
\ee
where $\theta$ is a phase of the superconducting order parameter and $\vA$ is a vector potential of the total magnetic field. Using the symmetry of the problem and integrating the expression above along a circular contour $R$, one obtains
\be
	\left. - \frac c {4\pi} \, \frac{\pa H}{\pa r} \right|_R
	=
	-
	\frac{c \phio} {8 \pi^2 R   \lambda^2} \,  \big( \ell +1 /\phio \, \oint_{R} \vA \cdot d \bm l \big),
\label{def of FO}
\ee
where the current density has been expressed in terms of the field derivative with the help of the Maxwell equation. The constant  $\ell$ characterizes the order parameter phase winding around the integration contour.

In applying the FO relation (\ref{def of FO}), one needs to bear in mind that the domain wall, being a phase defect, may possess a nonzero vorticity and, hence, along with vortices, contributes to the winding number $\ell$ for appropriate integration contours. For a circular contour the domain wall vorticity  is denoted as $\ell_{\rm dw}$. For a nonzero $\ldw$ not only the magnetic field but also the screening current experience a jump across the domain wall. This follows from writing down (\ref{def of FO}) for inner and outer boundaries of the domain wall. Taking into account  that the flux is a continuous function of the integration contour, one obtains
\be
	(\left. \pa H/ \pa r) \right|_{2+} - (\left. \pa H/ \pa r) \right|_{2-}= \frac{\phio}{2 \pi R_2 \lambda^2}\, \ell_{\rm dw},
	\label{field derivative across DW}
\ee
where $R_2$ is the radius of the circular domain wall.

Recalling that due to the presence of the chiral current the magnetic field itself experiences a jump across the domain wall and using (\ref{field derivative across DW}) and (\ref{London eqn. solutions}) we obtain an expression that relates field values on  boundaries 1,2 and 3:
\begin{eqnarray}
	H_{2-} a_{13} & = & H_{1+} a_{23} + H_{3-} a_{12} - \ti \, \tR_2^2 a_{12} (b_{23} - a_{23})
	\nonumber
	\\
	& +&
	\frac{\phio}{2\pi \lambda^2} \ell_{\rm dw} \, a_{12}a_{23},
	\label{bc on 2}
\end{eqnarray}
where $\tR \equiv R/\lambda$ and $b_{23}$ is defined by\footnote{Utilizing well-known relations between Bessel functions it is possible to use their alternative combinations  e.g.~those involving Bessel functions of zeroth and first order. In particular $b_{23} - a_{23} = \frac 2 {\tR_2} \big( I_0 (3) K_1(2) + K_0(3) I_1(2) \big)$. Our definitions follow those of Ref.~\onlinecite{Arutunian:1983}.}
\be
	b_{jk} \equiv K_2(j) I_0 (k) - I_2(j) K_0 (k).
	\label{def of b}
\ee
As a useful check of various identities one can consider limiting cases of domain wall 2 moving to either the inner or the outer surface of the cylinder: $2 \to 1$ or $ 2 \to 3$. For example in the limit $2 \to 1$ relation (\ref{bc on 2}) becomes $H_{2-} = H_{1+}$ which implies that in a very thin domain the field is uniform. In the limit $2 \to 3$ we have $H_{2-} = H_{3-} - 2 \ti $ or using eqns.~(\ref{field jumps}), $H_{2+} = H_{3-}$, thus reaching the same conclusion.\footnote{Notice that in the limit $R_j \to R_k$: $a_{jk} \to 0$ and $b_{jk} \to 2/\tR^2_j$. In general $a_{jk}=-a_{kj}$ but $b_{jk}$ cannot be expressed in terms of $b_{kj}$.} 

Another equation which relates $H_{1+}$ and $H_{2-}$ can be found by writing the FO quantization condition (\ref{def of FO}) for boundary 1. Taking into account that the flux through the hole is determined by the field $H_h$, which is related to $H_{1+}$ as $H_{1+}=H_h -\ti$, one obtains
\be
 	H_{1+} = - \frac {a_{12}}{b_{12}} \, \frac{\phio} {\pi R_1^2} \, \ell_s + \frac{2}{\tR_1^2 b_{12}} H_{2-} - \ti\,\frac {a_{12}}{b_{12}},
	\label{bc on 1}
\ee
where $\ell_s$ is the number of vortices trapped in the hole. Taking the $2\to 1$ limit in the above equation yields $H_{1+}=H_{2-}$ confirming the expectation that in a vanishingly thin domain the field is uniform.

Combining eqns.~(\ref{bc on 2}) and (\ref{bc on 1}) allows one to determine the field values in the hole $H_h$ and on the domain wall, $H_{2-}$, in terms of the applied field $H_a$. The field in the hole is given by the following expression:
\be
\begin{split}
	H_h & =
	-  \frac{a_{13}}{b_{13}} \, \frac {\phio}{\pi R_1^2}\, \ell_s -
	\frac{a_{23}}{b_{13}} \, \frac {\phio}{\pi R_1^2}\, \ell_{\rm dw} +
	 \frac{2}{\tR_1^2 b_{13}} H_{a}
	+
	\ti \, X_1,
	\\
	X_1 & \equiv 1- \tR_2^2 a_{12} - \frac{a_{13}}{b_{13}} \, (1- \tR_2^2 b_{12}) + \frac{2}{\tR_1^2 b_{13}} ( 1 - \tR_2^2 b_{23}),
	\label{Hh}
\end{split}
\ee
and the limiting values of the field-independent constant $X_1$ are equal to
\be
\begin{split}
	2 \to 3: \quad
	X_1 & = 1 - \frac 1 {b_{12}} \, (a_{12} + \frac 2 {\tR_1^2} ),
	\\
	2 \to 1: \quad 
	X_1 & = -1 + \frac 1 {b_{13}} \, (a_{13} + \frac 2 {\tR_1^2}).
	\label{X1 limits}
\end{split}
\ee
Since in the limits given above the domain structure degenerates to a single domain with positive ($2\to1$) or negative ($2\to 3$) chirality the constant $X_1$ changes sign as expected.

The field on the inner side of the domain wall is given by
\be
\begin{split}
	H_{2-}
	& =
	- \frac{a_{23}}{b_{13}} \, \frac \phio {\pi R_1^2} \, \ell_s - 
	\frac{b_{12} a_{23}}{b_{13}} \, \frac \phio {2 \pi \lambda^2} \, \ell_{\rm dw} + 
	\frac {b_{12}}{b_{13}} \, H_a + \ti X_2,
	\\
	X_2  & \equiv 
	-\frac{a_{23}}{b_{13}} \, (1- \tR_2^2 b_{12}) + \frac{b_{12}}{b_{13}}\, (1 - \tR_2^2 b_{23}).
	\label{H2}
\end{split}
\ee
In the limiting cases the field-independent constant $X_2$ reduces to
\be
\begin{split}
	2 \to 3: \quad
	X_2 & = -1
	\\
	2 \to 1: \quad
	X_2 &  = -2 + \frac 1 {b_{13}} \, (a_{13} + \frac 2 {\tR_1^2}).
	\label{X2 limits}
\end{split}
\ee
The first of these equations implies, in particular, that for a zero applied field $H_a=0$ and zero winding number $\ell_s=0$ the field at the inner side of the external boundary in the limit $2\to 3$ is equal to $-\ti$, as expected in this single-domain zero-field limit. The limit $2 \to 1$, however, deserves a more careful consideration. In this limit $H_{2-}$ should be related to the field in the hole as $H_{2-} = H_h - \ti$ and for the zero-applied field zero-winding number case the field in the hole can be read from eqn.~(\ref{X1 limits}). Comparing the second of eqns.~(\ref{X1 limits}) to the second of eqns.~(\ref{X2 limits}), one indeed recovers such a result.

Having expressed values of the field on all domain boundaries in terms of the applied field, the chiral current, and the winding numbers [eqns.~(\ref{Hh}) and (\ref{H2})] we thus find the field distribution in the entire system [eqns.~(\ref{London eqn. solutions}), (\ref{def of c}) and (\ref{def of a})]. 

To write  the Gibbs potential of the system requires knowledge of fluxes through areas limited by the domain boundaries. Although they can be found by the direct integration of (\ref{London eqn. solutions}) it is convenient to find fluxes surrounded by domain boundaries using the FO quantization condition. Doing so for the second domain boundary at $r=R_2 +0$ we get
\be
	\frac{a_{23} \phio}{2 \pi \lambda^2}\, (\ell_s + \ell_{\rm dw}+\Phi_2/\phio) = H_{3-} - H_{2+}\, \frac{R_2^2}{2 \lambda^2} \, (b_{23}-a_{23}),
\ee
which can be used to find $\Phi_2$. Alternatively, flux $\Phi_2$ can also be expressed in terms of fields $H_{2-}$ and $H_{1^+}$ which corresponds to writing FO for the inner part of domain wall 2 i.e.~for $r=R_2-0$.

Similarly, for $\Phi_3$ we obtain
\be
	\frac{a_{23} \phio}{2 \pi \lambda^2}\, (\ell_s + \ell_{\rm dw}+ \Phi_3/\phio) = -H_{2+} + H_{3-}\, \frac{R_3^2}{2 \lambda^2} \, (b_{32}+a_{23}).
\ee	
Notice that flux $\Phi_3$, i.e.~the total flux through the area limited by the outer boundary of the cylinder, determines its magnetic moment $M$.

Finally, flux $\Phi_1$ through the opening is determined by the field in the hole:
\be
	\Phi_1 = \pi R_1^2 H_h.
\ee

\subsection{Single-domain limit}

The results for a two-domain cylinder given earlier  can be used to find the field values in a single-domain limit. This is achieved by moving the domain wall located at $R_2$ either to the outer $R_2 \to R_3$ or to the inner $R_2 \to R_1$ boundary of the sample. Doing the former, one recovers a single-domain sample with a negative chirality, whereas in the latter case  a positive chirality sample with  $\ldw$ extra flux quanta trapped in the hole is obtained.

Taking the limit $R_2\to R_3$ the Gibbs potential of a single-domain chiral cylinder with a hole is given by the expression
\be
	8\pi g_{\text{s-d}} = - \phio \ell_s (H_h - H_a) - H_a \Phi_3 + \ti (\Phi_3 - \Phi_1),
\ee
where to agree with the previous notation the total flux through the cylinder is denoted as $\Phi_3$ and $\Phi_1$ is the flux through the hole. The field in the hole is given by
\begin{eqnarray}
	H_h  &=&
	-  \frac{a_{13}}{b_{13}} \, \frac {\phio}{\pi R_1^2}\, \ell_s 
	+
	 \frac{2 \lambda^2}{R_1^2 b_{13}} H_{a}
	 \nonumber
	 \\
	&+&
	\ti \, \big(-1 + (a_{13} + 2\lambda^2/R_1^2)/b_{13}\big).
\end{eqnarray}
and for the flux through the area limited by the outer boundary we obtain
\be
	\frac{a_{13} \phio}{2 \pi \lambda^2}\, (\ell_s + \Phi_3/\phio) = -(H_h +\ti) + (H_a+\ti )\, \frac{R_3^2}{2 \lambda^2} \, (b_{31}+a_{13})
\ee

Notice that it is  possible to obtain the single-domain results quoted above directly from the results of Ref.~\onlinecite{Arutunian:1983} for a nonchiral cylinder using the following replacements: $\ell_s \to \ell_s  - \pi R_1^2 \ti /\phio$, $H_a \to H_a + \ti$, and $H_h \to H_h +\ti$.

\section{Some useful expressions involving modified Bessel functions}

\label{appendix 2}

In this appendix we collect some useful results which involve modified Bessel functions. The unit of length is set to $\lambda$.

Given definitions (\ref{def of a and b}) of  $a_{ij}$ and $b_{ij}$ one can show that
\be
	K_0(R_1)I_1(R_2)+I_0(R_1) K_1(R_2) =\frac{R_2} 2 (b_{21}+a_{12})
\ee
and
\be
	a_{ij}=\frac{R_k^2} 2 (b_{ki} a_{kj} - b_{kj}a_{ki})
\ee 

Below we give limiting values of several parameters required to obtain the results of Section \ref{micro} from general formulas given in Appendix \ref{appendix 1}. 

For $R_3 \gg 1$ the following hold
\be
	a_{13} \to K_0(1) \frac{e^{R_3}}{\sqrt{2\pi R_3}}
	\big(1+\frac1{8R_3} + {\cal O}({R_3^{-2}}) \big)
\ee
\be
	b_{13} \to K_2(1) \frac{e^{R_3}}{\sqrt{2\pi R_3}} \big(1+\frac1{8R_3} + {\cal O}({R_3^{-2}}) \big)
\ee
\be
	b_{31} \to -K_0(1) \frac{e^{R_3}}{\sqrt{2\pi R_3}} 
	\big(1-\frac{15}{8R_3} + {\cal O}({R_3^{-2}})\big)
\ee



\end{document}